\documentclass{PoS}

\usepackage{epsfig}

\title{Numerical study of the mass spectrum in the 2D O(3) sigma model with a theta term}

\ShortTitle{Mass spectrum in the 2D O(3) non--linear sigma model with a theta term}

\author{\speaker{B. All\'es}\\
        INFN Sezione di Pisa, Pisa, Italy\\
        E-mail: \email{alles@df.unipi.it}}
\author{A. Papa\\
        Dipartimento di Fisica, Universit\`a
        della Calabria and INFN, Gruppo collegato di Cosenza\\
        Arcavacata di Rende, Cosenza, Italy\\
        E-mail: \email{papa@cs.infn.it}}

\abstract{It has been conjectured that the mass spectrum of the O(3)
non--linear $\sigma$ model with a $\theta$ term in 2 dimensions may
possess an excited state, which decays when $\theta$ is lowered from
$\pi$ below a critical value. Since the direct numerical investigation
of the model is prevented by a sign problem, we try to infer some
information on the mass spectrum at {\it real} $\theta$ by studying
the model at {\it imaginary} $\theta$ via analytic continuation.
A modified Swendsen--Wang cluster algorithm has been introduced to
simulate the model with the $\theta$ term.}

\FullConference{The XXV International Symposium on Lattice Field Theory\\
		 July 30 - August 4 2007\\
		 Regensburg, Germany}

\begin{document}

\section{Introduction}

Integrable Quantum Field Theories in 2 dimensions are successfully studied
by using the S--matrix approach both for massive~\cite{zamol1} and
massless theories~\cite{zamol2,zamol3}.
On the other hand the study of non--integrable models is
complicated because their scattering
amplitudes are generally non--elastic. These models are in fact characterized
by particle production, resonances, decay events, etc. along with the simple pole
structure that features integrable theories.

The O(3) non--linear $\sigma$ model in 2 dimensions with a $\theta$ term is a
well--known example of non--integrable theory. It is defined by the
action~\cite{poly}
\begin{equation}
 S=\frac{1}{2g} \int \hbox{d}^2x \left(\partial_\mu \vec{\phi}(x)\right)^2 -
   i \theta \int \hbox{d}^2x \,Q(x)\;,
\label{continuumS}
\end{equation}
where $g$ is the coupling constant, $\vec{\phi}(x)$ is a 3--component
unit vector and $Q(x)$ is the topological charge
density operator~\cite{belavin}  ($a,b,c$ are
group indices that run from 1 to 3 and $\mu,\nu$ are 2--dimensional spacetime indices)
\begin{equation}
 Q(x)=\frac{1}{8\pi} \epsilon^{\mu\nu} \epsilon_{abc} \phi^a(x) \partial_\mu \phi^b(x)
      \partial_\nu \phi^c(x)\; .
\label{continuumQ}
\end{equation}
The integration of $Q(x)$ throughout the whole 2--dimensional spacetime
yields the total topological charge $Q$ which takes on integer numbers
that reveal the winding of configurations over the sphere S$^2$.
Topological charge $(-)1$ configurations are called (anti)instantons.

When $\theta=0$ or $\pi$ the model is integrable and its spectrum is as follows. For
$\theta=0$ it contains a triplet of massive scalars whose mass has been
analytically calculated~\cite{hasenfratz} and numerically verified within a
2\%--3\% error~\cite{alles}. At $\theta=\pi$ it has been conjectured that
the theory becomes massless~\cite{zamol3,affleck} (for a numerical analysis
of the corresponding universality class see~\cite{bietenholz}).

The study of the evolution of the spectrum as $\theta$ moves from $\pi$
towards lower values is worthwhile. The model contains a triplet and a singlet
states whose masses ($m_T$ and $m_S$ respectively)
are proportional to $(\pi -\theta)^{2/3}$ for
$0<(\pi -\theta)\ll 1$~\cite{affleck2}. By using Form Factor Perturbation
Theory~\cite{muss1} it has been shown that the singlet is heavier than the
triplet~\cite{muss2} (see also Ref.~\cite{camposvenuti}). In particular
at $\theta\approx\pi$ the ratio of singlet to triplet masses is $m_S/m_T\approx\sqrt{3}$.

On the other hand it is known that at $\theta=0$ there is no lingering trace of
the singlet state. This fact suggests that the singlet mass may diverge as $\theta$
approaches zero, thus decoupling from the whole theory. Consequently, by continuity
arguments, it seems plausible that there exist a critical $\theta_c$ where
the singlet mass becomes exactly twice the triplet mass in such a way that for
all $\theta < \theta_c$ the singlet decays into states belonging to the
triplet.

\begin{figure}
\vspace{0.7cm}
\hspace{2cm}
\epsfig{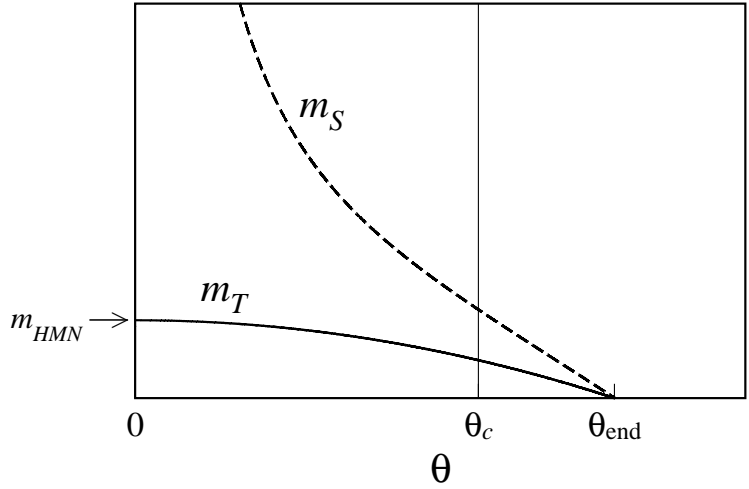}
\caption{Sketch of the hypothesized $\theta$ dependence for the singlet (dashed
line) and triplet (continuous line) masses in the O(3) non--linear
sigma model in 2 dimensions.
$\theta_c$ is the value where the
singlet mass becomes larger than twice the triplet mass.
$m_{HMN}\equiv m_T(\theta=0)$ is the mass calculated in Ref.~\cite{hasenfratz}.
$\theta_{\rm end}$ is the endpoint where the theory becomes massless. It has
been conjectured that $\theta_{\rm end}=\pi$.}
\label{Fig1}
\end{figure}

In Fig.~\ref{Fig1} a sketch representing the qualitative behaviour expected for the
$\theta$ dependence of the spectrum of the theory is shown.
We are currently making a Monte Carlo study of the model to examine some of the
salient features shown in this Figure. In the present progress report
we give clear numerical indications that $m_T(\theta)$
vanishes at the endpoint $\theta_{\rm end}=\pi$ (for an extensive account
see~\cite{alpapa}).

To this end we have prepared a code that simulates the 2--dimensional
O(3) sigma model with an {\it imaginary} $\theta$ term. After extracting
the triplet mass with this code, one can analytically continuate the
results to {\it real} values of $\theta$ thus obtaining a numerical
estimate of the lower curve in Fig.~\ref{Fig1}. We focus our attention on
the location of the endpoint $\theta_{\rm end}$ on this curve.
In section~2 we shall shortly
describe the updating algorithm used at imaginary $\theta$.
In section~3 the analytic continuation is performed and the
main results are exhibited.

\section{The Lattice implementation}

We have regularized the model in Eq.(\ref{continuumS}) by using the standard
lattice action
\begin{equation}
 S_L = A_L - i \theta_L Q_L \;, \qquad 
 A_L \equiv-\beta \sum_{x,\mu} \vec{\phi}(x)
 \cdot \vec{\phi}(x+\widehat{\mu}) \;,\qquad
 Q_L \equiv \sum_x Q_L(x)\; ,
\label{latticeS}
\end{equation}
where $\beta$ is the inverse bare lattice coupling constant for this
standard regularization and $\theta_L$ is the bare theta parameter. The corresponding
endpoint and critical point will be called $\theta_{L,{\rm end}}$ and
$\theta_{L,c}$. The lattice operator for the density of topological
charge is defined as in~\cite{papa1},
\begin{equation}
Q_L(x)=\frac{1}{32\pi} \epsilon^{\mu\nu} \epsilon_{abc} \phi^a(x)
       \Big(\phi^b(x + \widehat{\mu}) - \phi^b(x - \widehat{\mu})\Big)
       \Big(\phi^c(x + \widehat{\nu}) - \phi^c(x - \widehat{\nu})\Big)\; .
\label{latticeQ}
\end{equation}
The coordinate sites are labelled by $x\equiv(x_1,\, x_2)$.
The total topological charge $Q_L$
obtained from the sum over the whole lattice volume of the above expression
does not yield integer numbers on {\it single} configurations. This is not an
inconvenience inasmuch as Quantum Field Theory deals with quantum averages
of renormalized operators over {\it many} configurations. Thus the above
definition for the density of topological charge~(\ref{latticeQ}) must be
renormalized, the renormalized charge being $Q$
\begin{equation}
Q_L = Z_Q \, Q \;,
\label{ZQ}
\end{equation}
and $Z_Q$ being the corresponding renormalization constant. It can be
calculated either by perturbation theory~\cite{pisa1} or by a
non--perturbative numerical method~\cite{pisa2}.
The meaning of this constant is simple: it accounts for the average
over quantum fluctuations in such a way
that $Q$ yields an integer value. Actually
this last observation is the basis for the above--mentioned
non--perturbative numerical method (see below).

The reason to choose the expression~(\ref{latticeQ}) to be used in the
lattice action is that, as we shall see later, it allows the introduction
of a variant of the cluster algorithm in the presence of the
$\theta$ term. There exist other lattice regularizations of the operator
$Q(x)$ that do not require the computation of $Z_Q$ (since it is exactly~1
for all $\beta$); however it is not easy to introduce a fast cluster
algorithm for them.

Our interest concerns the spectrum of the theory
for varying $\theta$. This parameter is related to the corresponding
bare $\theta_L$ by the expression $\theta=Z_Q\,\theta_L$. Therefore
the value of the bare theta parameter $\theta_{L,{\rm end}}$ where the mass
vanishes is conjectured to be $\theta_{L,{\rm end}}=\pi/Z_Q$.
This expression is a function of $\beta$ since $Z_Q$ in general depends
on the coupling constant.

We have run our Monte Carlo simulations at imaginary values of the theta
parameter, $\theta_L=+i\vartheta_L$ ($\vartheta_L\in{{\rm I}\kern-.12em{\rm R}}$)
in order to avoid the sign problem. Let us briefly introduce the new
cluster algorithm for imaginary $\theta_L$.

The first part of an updating step with the Wolff algorithm~\cite{swendsen}
for the standard O(3) sigma model without a $\theta$ term consists in choosing
a random unit vector $\vec{r}$ in such a way that every dynamical
field can be split in a component parallel to $\vec{r}$ and the rest,
$\vec{\phi}(x)=\left(\vec{\phi}(x)\cdot\vec{r}\right)\vec{r} + \vec{\phi}_\bot(x)$,
where $\vec{\phi}_\bot(x)$ denotes the part of $\vec{\phi}(x)$ orthogonal
to $\vec{r}$. Then the signs of $\left(\vec{\phi}(x)\cdot\vec{r}\right)$
for all $x$ are updated \`a la Swendsen--Wang as in the Ising model.

By introducing the above separation for $\vec{\phi}(x)$ in the
expression~(\ref{latticeQ}) we can rewrite it as
\begin{eqnarray}
Q_L(x)&=&\frac{1}{16\pi} \Big\{
  \left(\vec{\phi}(x)\cdot\vec{r}\right)
      \left(d_{1,2} + d_{-1,-2} + d_{2,-1} + d_{-2,1}\right) + \nonumber \\
  &&\qquad\;\;\;\left(\vec{\phi}(x+\widehat{1}\,)\cdot\vec{r}\right)
      \left(d_{0,-2} - d_{0,2}\right) +
  \left(\vec{\phi}(x-\widehat{1}\,)\cdot\vec{r}\right)
      \left(d_{0,2} - d_{0,-2}\right) + \nonumber \\
  &&\qquad\;\;\;\left(\vec{\phi}(x+\widehat{2}\,)\cdot\vec{r}\right)
      \left(d_{0,1} - d_{0,-1}\right) +
  \left(\vec{\phi}(x-\widehat{2}\,)\cdot\vec{r}\right)
      \left(d_{0,-1} - d_{0,1}\right) \Big\}\;,
\label{latticeQ2}
\end{eqnarray}
where $x\pm\widehat{1}$ means the site at the position one step forward
(backward) in the direction ``1'' starting from site $x$
and the notation $d_{i,j}$ stands for the $3\times 3$ determinant
(the three components for each vector are unfold along the rows)
\begin{equation}
d_{i,j}\equiv \det\pmatrix{\vec{r}\cr
             \vec{\phi}(x+\widehat{i}\,\;)\cr
             \vec{\phi}(x+\widehat{j}\,\;)\cr}\; .
\label{definitiond}
\end{equation}
In this fashion the theory at each updating step looks like an Ising model
in the presence of an external local magnetic field $h(x)$ because the
$\theta$ term in Eq.(\ref{latticeQ2}) is linear in
$\left(\vec{\phi}\cdot\vec{r}\right)$.
One can readily derive that the magnetic field at site $x$ is
\begin{eqnarray}
 h(x)=-\frac{\vartheta_L}{16\pi}\vert\vec{\phi}(x)\cdot\vec{r}\vert &&
\Big(d_{1,2} + d_{-1,-2} + d_{2,-1} + d_{-2,1} + d_{-1,-1-2} + d_{-1+2,-1} +
\nonumber \\
     && \;\;\,d_{1,1+2} + d_{1-2,1} + d_{2,2-1} + d_{2+1,2} + d_{-2,-2+1} + d_{-2-1,-2}\Big)\;.
\label{fieldmag}
\end{eqnarray}
$d_{i+k,j}$ (and analogous terms in~(\ref{fieldmag}))
are the straightforward generalization of the above
definition~(\ref{definitiond}) when the site is obtained by shifting
two steps ($\;\widehat{i}$ plus $\widehat{k}\;$) from the original
position~$x$.

We have converted the original theory in an Ising model
in the presence of a local external magnetic field (which changes at each
updating step and therefore must be recalculated at each step). In the
literature there are two algorithms expressly introduced to update
Ising models in the bosom of magnetic fields: the Lauwers--Rittenberg~\cite{lauwers}
and the Wang~\cite{wang} methods.
We tested the performances of both algorithms, compared their decorrelation
times and decided for the latter. The cluster construction
was tackled by the Hoshen--Kopelman algorithm~\cite{hoshen}.

\section{Analytic continuation and determination of $\theta_{\rm end}$}
\label{analytical}

Operators which couple the vacuum with the singlet or triplet states
can contain an arbitrary number of fundamental fields since the model
is not parity invariant for $\theta\not=0$. We only present the results
for the triplet particle. The following operators were used
\begin{equation}
{\overrightarrow{\cal O}}_1(x)\equiv\vec{\phi}(x)\,,\qquad\qquad
{\overrightarrow{\cal O}}_2(x)\equiv-i\vec{\phi}(x)\times\vec{\phi}(x+\widehat{1}\,)\,.
\end{equation}
Then we calculated the related wall operators by averaging over the
$x_1$ coordinate (as usual $L$ is the lattice size),
${\overrightarrow{\cal W}}_i(x_2)\equiv \frac{1}{L}\sum_{x_1}
{\overrightarrow{\cal O}}_i(x)$ for $i=1,2$.

\begin{figure}
\vspace{0.7cm}
\hspace{2cm}
\epsfig{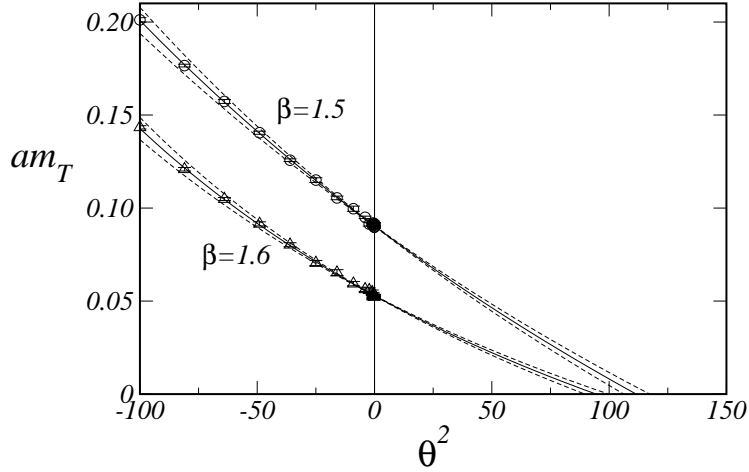}
\caption{Behaviour of the triplet mass (in units of the lattice spacing)
as a function of $\theta^2$. Circles ($\beta=1.5$) and triangles ($\beta=1.6$)
are the data from the simulation at imaginary $\theta$ ($\theta^2<0$). Each
continuous line is the result of the analytic continuation described in the text
and the dashed lines enclose the boundary of its error.}
\label{Fig2}
\end{figure}

\begin{figure}
\vspace{0.7cm}
\hspace{2cm}
\epsfig{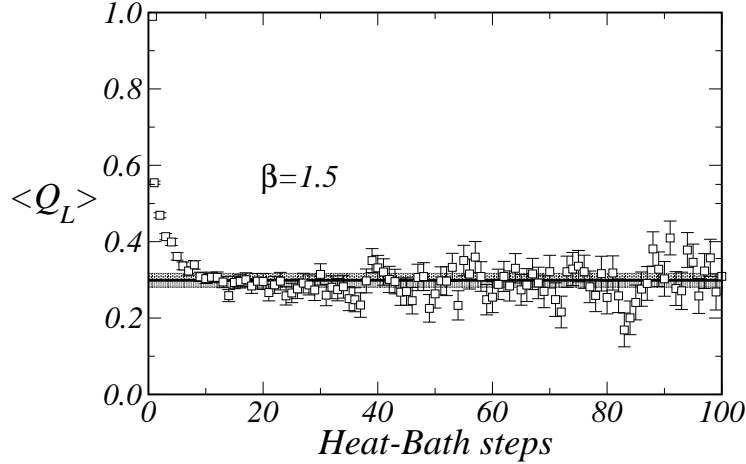}
\caption{Data for $\langle Q_L\rangle$ start at $+1$
at the 0--th Heat--Bath step and
then they go down until reaching a plateau. The horizontal line and grey
band are the value and error respectively of $Z_Q(\beta=1.5)$.}
\label{Fig3}
\end{figure}

To single out the correct parity mixture for the physical particle and
to clean the signal from possible excited states, we extracted the triplet
mass $m_T$ by using the variational method of Ref.~\cite{kronfeld} where
the mass is obtained from the largest eigenvalue of the correlation matrix
$\langle {\overrightarrow{\cal W}}_i(x_2){\overrightarrow{\cal
    W}}_j(0)\rangle - \langle{\overrightarrow{\cal W}}_i\rangle
\langle{\overrightarrow{\cal W}}_j\rangle$.
In Fig.~\ref{Fig2} the results for the triplet mass are shown for two
values of $\beta$. The analytic continuations in this figure
were done by using a ratio of degree~2
polynomials; analogous results are obtained from other functional
forms (we checked this statement by using degree~4 and~6 polynomials).
The endpoint (where the mass vanishes) equals
$\left(\theta_{L,{\rm end}}\right)^2=(\theta_{\rm end}/Z_Q(\beta))^2$
where $\theta_{\rm end}$ is conjectured to be equal to $\pi$.
On the other hand the renormalization constant $Z_Q$ can be evaluated
with the non--perturbative
numerical method of Ref.~\cite{pisa2}. An example of
such an evaluation is shown in Fig.~\ref{Fig3}. Summed up briefly: a classical
instanton (with topological charge $+1$) is put by hand on the lattice and
then it is heated by applying
100 updating steps (we used Heat--Bath steps on the conventional O(3)
non--linear sigma model without a $\theta$ term since $Z_Q$ cannot depend
on $\theta$). The value of $Q_L$ is
measured while the continuum topological charge $Q$ is continuously
monitored (by 6 cooling steps after every Heat--Bath updating)
to be sure that the background charge is not changed. This procedure
was repeated $10^4$ times for both values of $\beta$. The average of
$Q_L$ on configurations that lie in the topological sector $+1$ yields $Z_Q$.

\begin{table}[hbtp]
{\centerline{Table 1}}
\vskip 2mm
\setlength{\tabcolsep}{1.88pc}
\centering
\begin{tabular}{cclllll}
\hline
\hline
$\beta$ & $L$ &  $\left(\theta_{L,{\rm end}}\right)^2$ 
        & $Z_Q$  & $\theta_{\rm end}$ \\
\hline
1.5     & 120   & 111(5) & 0.299(18) & 3.15(20) \\
\hline
1.6     & 180   & 94(5)  & 0.313(12) & 3.03(14) \\
\hline
\hline
\end{tabular}
\end{table}

Then, once we know $Z_Q$ and $\left(\theta_{L,{\rm end}}\right)^2$ the
value of $\theta_{\rm end}$ can be extracted. The results are shown
in Table~1. The lattice sizes were chosen large
enough to guarantee the absence of relevant finite size effects (we imposed
$L/\xi\equiv L\cdot am_T\ge 10$).
$2\cdot 10^5$ propagators were measured on as many
independent configurations for all values of $\beta$ and $\theta_L$. The results
of the last column are in fair agreement with the conjecture that predicts
$\theta_{\rm end}=\pi$. By averaging over the results for both $\beta$
(assuming gaussian errors) we obtain
that the endpoint for the triplet mass is $\theta_{\rm end}=3.07(11)$.

\section{Conclusions and outlook}

We are studying the spectrum of the 2D O(3) non--linear sigma model with a
$\theta$ term as a function of this parameter. The model contains a
triplet and a singlet states whose mass depend on $\theta$ as shown in
the sketch of Fig.~\ref{Fig1}. In the present progress report we
have given clear evidence that the triplet mass indeed vanishes when
$\theta$ becomes $\pi$. To do so, we have simulated the model at imaginary $\theta$
and then extrapolated the results to real $\theta$. The extrapolation always
indicate that the mass tends to vanish at an endpoint. Our calculation
yields $\theta_{\rm end}=3.07(11)$ in good agreement with the conjectured
prediction $\theta_{\rm end}=\pi$.

A new fast cluster algorithm was purposely introduced to simulate the theory at
imaginary $\theta$.

We are planning to improve the statistics by studying the model at other
values of $\beta$ and other lattice regularizations of the
topological charge as well as by increasing the precision in the evaluation
of $Z_Q$ which is the largest source of error in the calculation of
$\theta_{\rm end}$ (see Ref.~\cite{alpapa}).
We will also extend the analysis to the singlet mass
and the determination of $\theta_c$.

We emphasize the good performance of the analytic continuation in our
study. It is important for that to have got Monte Carlo data within a wide
range of values of $\theta_L$, ($\vartheta_L\equiv-i\theta_L\in[0,10]$).

\end{document}